\documentclass[a4paper]{article}
\usepackage[affil-it]{authblk}
\usepackage[english]{babel}
\usepackage{tikz}
\usepackage{subcaption}
\usepackage{amsmath}
\usepackage{amssymb}
\usepackage{overpic}
\usepackage{float}
\usepackage{placeins}
\usepackage{xurl}

\usetikzlibrary{calc}

\usepackage[margin=1.in]{geometry}

\definecolor{accentBlue}{RGB}{0, 114, 178}
\definecolor{accentOrange}{RGB}{213, 94, 0}
\definecolor{accentGreen}{RGB}{0, 158, 115}

\newcommand{\bs}[1]{\boldsymbol{#1}}

\title{Adversarial robustness of a U-Net-based model observer for CT protocol optimization}

\vspace{2cm}

\author[1]{Filippo Maria Balli}

\author[2]{Giorgia Stendardo}

\author[3,4]{Sandra Doria}

\author[1]{Michele Ginolfi}

\author[1]{Alessio Gnerucci}

\author[5]{Diego Sona}

\author[6]{Adriana Taddeucci}

\author[1]{Cesare Gori}

\author[2]{Evaristo Cisbani}

\affil[1]{Department of Physics and Astronomy, 
  Universit\`{a} degli Studi di Firenze (UNIFI),
  Via Giovanni Sansone 1,
  Sesto Fiorentino, 50019 -  Florence - Italy}

\affil[2]{Istituto Superiore di Sanit\`{a}, Centro Nazionale Intelligenza Artificiale e Tecnologie Innovative per la Salute (ISS-IATIS),
  Viale Regina Elena 299 - 00161 - Rome - Italy}

\affil[3]{Institute of Chemistry of Organometallic Compounds, National Research Council (ICCOM-CNR),
  Via Madonna del Piano 10,
  Sesto Fiorentino, 50019 - Florence - Italy}

\affil[4]{European Laboratory For Non Linear Spectroscopy (LENS), Universit\`{a} degli Studi di Firenze (UNIFI),
  Via Nello Carrara 1, Sesto Fiorentino,
  50019 - Florence - Italy}

\affil[5]{Fondazione Bruno Kessler (FBK),
  Via Sommarive 18, Povo,
  38123 - Trento - Italy}

\affil[6]{Ospedale Santa Maria Annunziata, Azienda USL Toscana Centro, Italy,
  Via Antella 58, Bagno a Ripoli,
  50012 - Florence - Italy}

\begin{document}

\maketitle

\begin{abstract}
Artificial intelligence is increasingly used in medical imaging, yet its robustness to input perturbations remains a critical concern for a wide clinical adoption. To this end, we used adversarial examples to systematically probe vulnerabilities of a U-Net-based model observer for computed tomography protocol optimization, performing detection and localization of low-contrast objects in a phantom dataset.
  Adversarial attacks were generated using both gradient-based and optimization-based white-box methods. Fast gradient perturbations produced high misclassification rates, reaching up to 75\% at intermediate perturbation levels while remaining visually imperceptible. Localization was more robust, with success rates of about 25\% for small perturbations and 42\% at moderate levels. In contrast, optimization-based attack achieved success rates close to 50\% for both tasks.
  To mitigate these vulnerabilities, dynamic adversarial training was implemented. This reduced the success rate of optimization-based attacks to 7\% for classification and 13\% when including localization-specific training, demonstrating a substantial robustness improvement without compromising task performances, confirmed by localization receiver operating characteristic analysis.
  To further interpret model behavior, radiomic texture analysis was performed on original and adversarial images. While most global image statistics remain stable, specific texture-related features exhibit consistent changes in successful attacks, highlighting the model's sensitivity to subtle local intensity patterns.
  Overall, adversarial training improves robustness without degrading performance, while radiomic analysis reveals interpretable links between texture alterations and prediction failures, supporting more reliable and explainable AI systems for medical imaging.
\end{abstract}

\section{Introduction}

Artificial intelligence (AI) is rapidly emerging as a transformative technology across numerous domains, including healthcare, due to the capacity of deep learning (DL) to model complex statistical systems and analyze large-scale datasets. For clinical adoption, however, data-driven DL tools require rigorous quantitative validation across experimental contexts and methodological variations. Two key properties warrant particular attention: robustness and generalizability. Robustness refers to an algorithm's resilience to deviations from underlying assumptions, specifically its ability to maintain performance under perturbations or noise in input data. Generalizability describes the algorithm's capacity to perform effectively on previously unseen data, such as datasets from different populations or production protocols. These properties are closely interrelated and intertwined with explainability \cite{freiesleben2023}.

Definitions of robustness vary widely. A broad, generic definition provided by ISO/IEC TR 24029-1:2021 describes it as \cite{ISOIEC24029-1:2021}: \emph{The ability of an AI system to maintain its level of performance under any circumstances}. In practice, validation under such broad conditions is unfeasible. A more operational definition was proposed by \cite{braiek2025machine}: \emph{When deployed in a production environment, a ML model is considered robust if variations of input data, as specified by a domain of potential changes, do not degrade the model's predictive performance below the permitted tolerance level.} This emphasizes the need to define an input data acceptance domain, assessing robustness against data shifts likely to occur in real-world scenarios.

One approach to quantify robustness involves adversarial examples (AEs) \cite{serban2018adversarial, Pedraza2021GeneralizationRobustness} and the related notion of universal adversarial perturbations \cite{moosavidezfooli2017universal, hirano21}, which exploit specific vulnerabilities of deep neural networks, such as evasion \cite{biggio2013evasion} or poisoning attacks. These methods enable the identification of input regions that maintain a desired performance level, indirectly informing generalizability. At the same time, adversarial examples are closely related to counterfactual explanations \cite{wachter2018counterfactual, freiesleben2022relation, molnar2025}---both involve input modifications that change a model prediction, though they differ in their objectives, constraints, and interpretability requirements---and can therefore also enhance interpretability by revealing conditions under which a model fails or succeeds. These examples are inputs deliberately designed to mislead deep learning models via imperceptible perturbations. Most deep learning models exhibit significant performance degradation when exposed to adversarial attacks, which can be mitigated through strategies such as re-training or transfer learning. In the medical imaging domain, deep learning models can rely on spurious features or shortcuts \cite{geirhos2020shortcut, degrave2021shortcuts}, and are specifically vulnerable to adversarial attacks that raise concerns for clinical reliability \cite{finlayson2019medical, paschali2018generalizability, dong2024survey}.

In this work, we exploit AEs on a case study involving an AI-based model observer (MO), previously proposed by \cite{valeri2023unet}, with the ultimate goal of evaluating and improving robustness. A model observer in medical imaging is designed to assess image quality by quantifying performance according to a predefined figure-of-merit (FOM) for a specific clinical task \cite{barrett1993modelobservers, barrett2004foundations, thno5138}. In \cite{valeri2023unet}, the authors developed and trained a U-Net-based MO to replicate human observer performance in confidence score prediction---implemented as a classification task---and localization of low-contrast objects in computed tomography (CT) scans obtained from a reference phantom. The clinical application focuses on optimizing CT protocols in terms of patient radiation dose, in line with the \emph{ALARA} principle \cite{ICRP1977} and relevant international and national regulations \cite{Euratom2013}.

We constructed various adversarial attacks targeting the MO algorithm, specifically designed to challenge both classification and localization tasks. Success rates of these attacks were evaluated across different perturbation levels, enabling identification of vulnerabilities and robustness breaks in the algorithm.

Finally, we performed radiomic analysis on original and adversarially perturbed CT images to identify features affecting the model predictions. Radiomic features---capturing intensity, shape, and texture patterns---link quantitative image descriptors to clinical data and are used in oncology for tumor characterization, prognosis, and treatment planning \cite{aerts2014quantitative, Gillies2016Radiomics, Mayerhoefer2020IntroductionRadiomics, vanTimmeren2020RadiomicsHowTo}. This analysis helps separate perturbation effects: uninformative features the model wrongly relies on and subtle features containing exploitable information invisible to humans \cite{ilyas2019features, buckner2020}. By comparing features from original and adversarial images, we reveal characteristics driving misclassification and enhance model explainability.

\begin{figure}
  \centering
  \resizebox{\columnwidth}{!}{
  \input{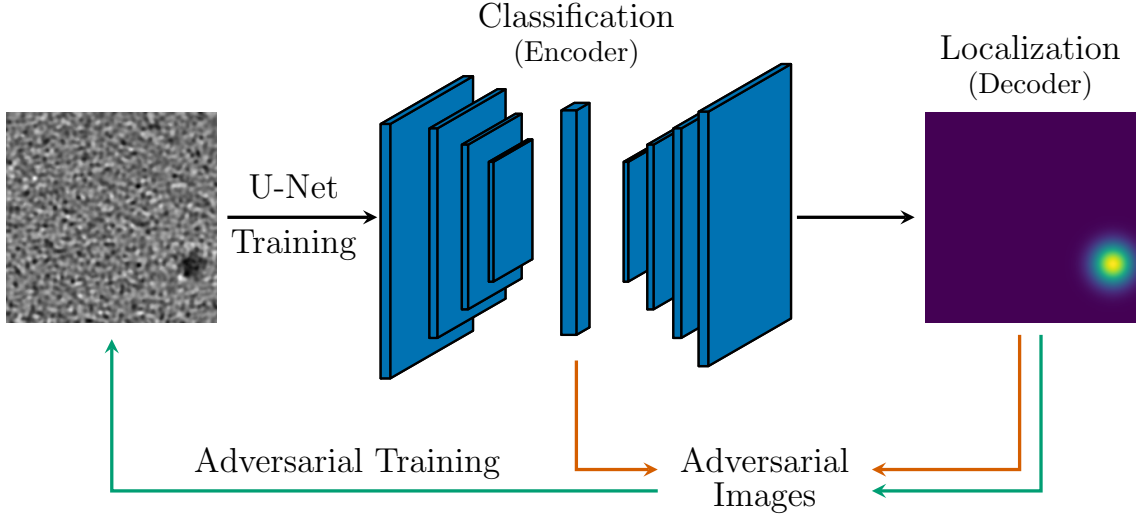}}
  \caption{Schematic representation of the analysis workflow: the U-Net encoder performs the classification task while its decoder performs insert localization as described in the text; both are considered for the adversarial attack, analysis and defense (across adversarial training).}
  \label{fig3.2_workflow}
\end{figure}

To improve MO robustness, we implemented different adversarial defense strategies across classification and localization tasks, identifying dynamic generation of adversarial examples during training as the most effective approach. Finally, the model's robustness and task performance were re-evaluated after adversarial training. The workflow of the adopted methodology is illustrated in Fig.~\ref{fig3.2_workflow}.

\section{Materials and methods}

\subsection{U-Net-based model observer}
\label{MOsubsection}

In this section, we briefly review the main features of the CT dataset and AI-based MO algorithm developed by \cite{valeri2023unet}, including its training methodology and details on its U-Net architecture.

\begin{figure}
  \centering
  \setlength{\tabcolsep}{2pt}
  \renewcommand{\arraystretch}{1.1}
    \begin{tabular}{cccc}
    & \footnotesize CTDI = 4.4 mGy
    & \footnotesize CTDI = 7.8 mGy
    & \footnotesize CTDI = 10.2 mGy \\
    \raisebox{6.5ex}{\rotatebox[origin=c]{90}{\footnotesize Diameter = 7 mm}} &
    \includegraphics[width=0.28\columnwidth]{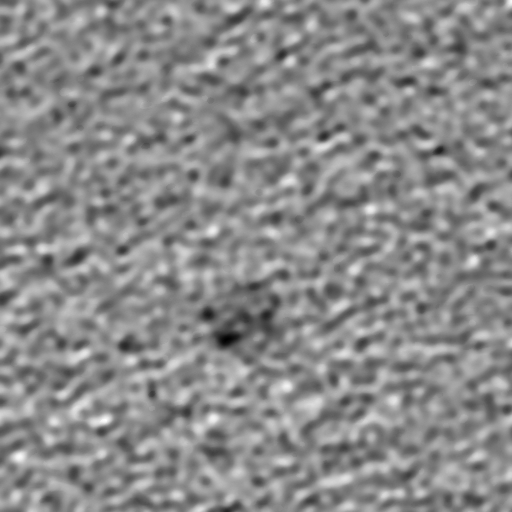}                &
    \includegraphics[width=0.28\columnwidth]{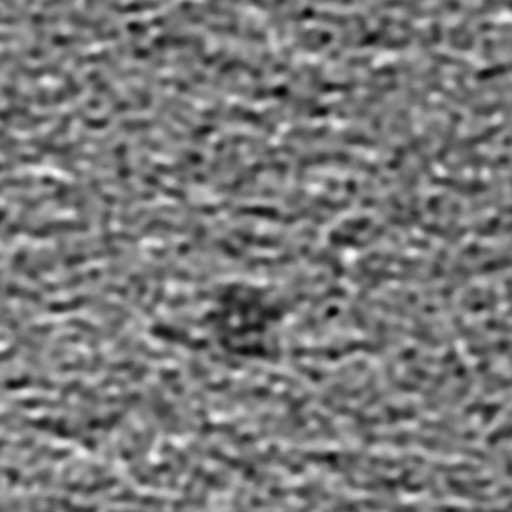}                &
    \includegraphics[width=0.28\columnwidth]{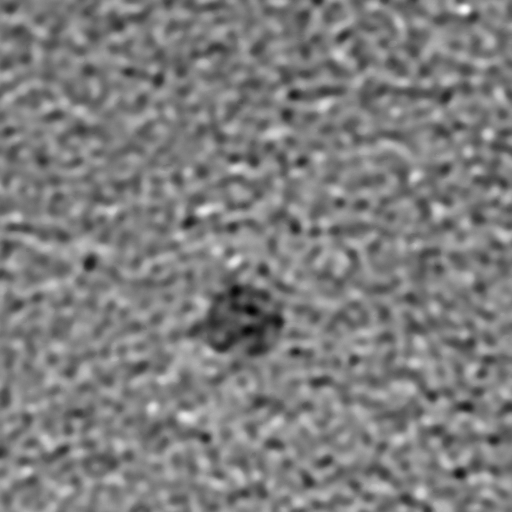}                                                        \\
    \raisebox{6.5ex}{\rotatebox[origin=c]{90}{\footnotesize Diameter = 4 mm}} &
    \includegraphics[width=0.28\columnwidth]{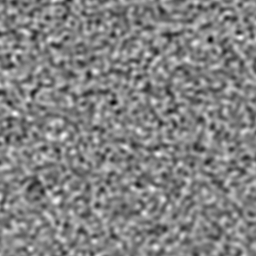}            &
    \includegraphics[width=0.28\columnwidth]{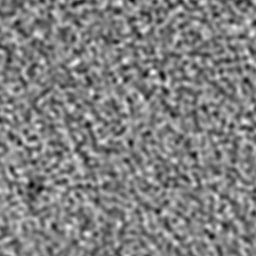}            &
    \includegraphics[width=0.28\columnwidth]{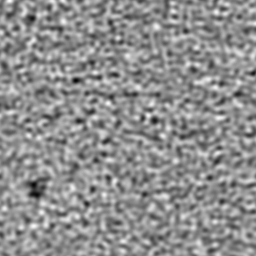}                                                    \\
  \end{tabular}
  \caption{Examples of reconstructed images from phantom CT dataset used in \cite{valeri2023unet} and \cite{doria2021addressing}.}
  \label{dosesfig}
\end{figure}

The dataset used to train, validate, and test the MO algorithm is a subset of a larger dataset described in a precedent work by \cite{doria2021addressing}.
It comprises CT images of a custom polymethyl methacrylate (PMMA) phantom with 10 cylindrical inserts of five different diameters ($3$~mm to $7$~mm) and two contrast levels ($45$~HU and $55$~HU). The phantom consists of three ellipsoidal blocks, two containing inserts and one homogeneous block providing background images. CT acquisitions were performed using a 128-slice CT scanner (Somatom Definition Flash, Siemens Healthcare) and selecting the standard oncological protocol for abdomen, at eight different CT dose index (CTDI) settings ($4.4$~mGy to $10.2$~mGy). Both filtered back projection (FBP) and iterative reconstruction (IR) techniques were used to reconstruct the acquired dataset. The CT image reconstruction field of view (RFoV) was set to $5$~cm$^2$ ($512 \times 512$ pixels per image, then reduced to  $256 \times 256$ pixels to optimize computational resources) to produce reconstructed images containing a single insert each. Similarly, images without inserts were reconstructed and added to the dataset. To further increase dataset variability, data augmentation techniques such as rotations and flips were applied to all images. From this extensive dataset, 30,000 images were selected for analysis. For more details on image acquisition, including the distribution of images based on insert diameter and contrast, refer to \cite{doria2021addressing}. Figure~\ref{dosesfig} shows selected reconstructed images at different CTDI and containing inserts of different sizes.

To generate labels for training the MOs, a multiclass ranking approach was employed. Human observers assessed each image on a 4-point scale from 0 (object not present) to 3 (object certainly present), reflecting their confidence in the recognition of inserts. Additionally,  for each image classified from 1 to 3, they pointed at the center of the identified potential insert providing the corresponding coordinates. These information were used as labels for MO training and testing \cite{valeri2023unet}.

The MO task in \cite{valeri2023unet} implemented a customized U-Net-based convolutional neural network (CNN) trained to perform both object localization and confidence score prediction. The U-Net, a CNN based on an autoencoder architecture (see Fig.~\ref{fig3.2_workflow}), is well-documented for its effectiveness in medical image segmentation and localization tasks \cite{ronneberger2015unet, Siddique2021UNetReview}.

In the MO implementation, the U-Net architecture consisted of nine layers with four skip connections \cite{ronneberger2015unet}. The encoder output is passed to a dense layer with a sigmoid activation, producing a scalar confidence score between 0 and 1. This continuous score is then discretized into four classes (0, 1, 2, 3) for evaluation, while the model is trained using a mean squared error loss on the normalized score. The decoder block focuses on localization, employing the autoencoder structure to generate heatmaps for position estimation \cite{Newell2016StackedHourglass, Payer2019HeatmapLandmark}. The model was optimized using a composite loss function:

\begin{equation}\label{TotalLoss_MO}
  L_{\text{U-Net}} = \lambda_{\text{MO}}\,L_{\text{MO}} + \lambda_{\text{LOC}}\, L_{\text{LOC}} + \lambda_{\text{KLD}}\,L_{\text{KLD}}
\end{equation}
where $L_{\text{MO}}$ is a mean squared error (MSE) loss between the predicted and target confidence scores, $L_{\text{LOC}}$ is a MSE loss between predicted and ground-truth coordinates, and $L_{\text{KLD}}$ is a Kullback-Leibler divergence (KLD) loss enforcing alignment between predicted and reference Gaussian maps. The loss weights $\lambda_{\text{MO}}, \lambda_{\text{LOC}}$ and $ \lambda_{\text{KLD}}$ were optimized during training. For further details on architecture parameters, optimization settings, and training protocol, refer to \cite{valeri2023unet}.

The results demonstrated that the combined classification and localization predictions from the CNN achieved excellent overall performance, measured in terms of Receiver-Operator-Characteristic (ROC) analysis, demonstrating the potential to employ the proposed methodology for CT protocols optimization.

\subsection{Adversarial examples}
Adversarial examples are inputs intentionally crafted to mislead deep learning models by introducing small, often imperceptible, modifications to otherwise correctly classified data. Since their first characterization by \cite{Szegedy2014Intriguing}, such examples have been shown to reliably induce misclassification, with success rates---defined as the fraction of perturbed inputs that cause incorrect predictions---approaching 100\% even in state-of-the-art architectures \cite{goodfellow2014explaining,Kurakin2016Adversarial}.
Conceptually, adversarial perturbations operate by shifting data points within the high-dimensional feature space from the decision region of the true class into that of another, leading to erroneous predictions.

In the following, we formalize the notation adopted for adversarial analysis. Consider a multi-task model represented by a function $f$ performing both classification and localization, consistent with the CNN architecture introduced in Sec.~\ref{MOsubsection}:
\begin{equation}
  (\ell, \boldsymbol{y}) = f(\boldsymbol{x}, \boldsymbol{\theta})
\end{equation}
where the input $\boldsymbol{x}$ is a vector of $n$ pixel intensities, $\boldsymbol{y} = (y_1, y_2)$ denotes the predicted coordinates for object localization, and $\ell$ is the predicted class label (score prediction). $\bs{\theta}$ is the set of trainable parameters adjusted during training; we will omit them in the following.

An adversarial example is obtained by applying the minimal perturbation $\boldsymbol{\eta}$ to a clean input $\boldsymbol{x}$ such that the model produces an incorrect prediction on the target task(s):
\begin{equation}
  \label{advi_definition}
  \min_{\boldsymbol{\eta}} \; \lVert \boldsymbol{\eta} \rVert_p
  \quad \text{s.t.} \quad f(\boldsymbol{x}) \neq f(\boldsymbol{x}_{\text{adv}})
\end{equation}
where $\boldsymbol{x}_{\text{adv}} = \boldsymbol{x} + \boldsymbol{\eta}$ denotes the adversarial example, and $\lVert \boldsymbol{\eta} \rVert_p$ is the $L_p$-norm of the perturbation.\footnote{For a perturbation vector $\boldsymbol{\eta} = (\eta_1, \dots, \eta_n)$, the $L_p$-norm is defined as $\lVert \boldsymbol{\eta} \rVert_p = \left( \sum_{i=1}^{n} |\eta_i|^p \right)^{1/p}$ for $p \geq 1$. Common choices in adversarial analysis are $p=2$, corresponding to the Euclidean norm, and $p=\infty$, which measures the maximum absolute perturbation over all pixels.}
Finding the exact minimal perturbation is generally challenging due to the non-linear and high-dimensional nature of modern deep networks \cite{Szegedy2014Intriguing}. As a result, adversarial examples are often generated by constraining the perturbation magnitude within a fixed bound \cite{goodfellow2014explaining},
\[
  \lVert \boldsymbol{\eta} \rVert_p \leq \delta,
\]
while still enforcing a change in the model's prediction. In many adversarial applications, it is further required that the perturbation remains imperceptible to human observers. Defining the human-perceived classification and localization outputs as
\begin{equation}
  (\ell_{\text{hum}}, \boldsymbol{y}_{\text{hum}}) = h(\boldsymbol{x})
\end{equation}
this condition can be expressed as the invariance of $h(\boldsymbol{x})$ under the perturbation introduced in \eqref{advi_definition}, namely $h(\boldsymbol{x}) = h(\boldsymbol{x}_{\text{adv}})$. This requirement ensures that the adversarial perturbation does not alter the perceptual or semantic content of the input, while still inducing an incorrect prediction by the model.

\subsubsection{Implementation of adversarial attacks on U-Net-based model observer}\label{sectionAttacks}

Numerous algorithms for generating adversarial images have been developed and are well-documented in the literature \cite{Xu2020AdversarialReview, Akhtar2018ThreatSurvey}. The adversarial attacks used in this work are \emph{white-box}, meaning that internal structure and parameters of the attacked model are known to the attacker.
All attacks in this study are \emph{untargeted}, intended only to induce an incorrect prediction rather than to force a particular target class. While \emph{targeted} attacks can offer insight into class-specific decision boundaries, they are outside the scope of this work, oriented to the evaluation of MO general robustness and prediction stability. In the following, two representative white-box attacks are considered: the fast gradient method (FGM) and the Carlini-Wagner $L_2$ attack (CW-$L_2$), briefly described in this section.

The \emph{fast gradient method} attack used here corresponds to the sign variant introduced by \cite{goodfellow2014explaining} as the \emph{fast gradient sign method (FGSM)}: a fast and effective algorithm for generating adversarial images. It operates on the assumption that small perturbations, applied in the direction of the gradient of the loss function $L(\boldsymbol{x},\ell)$ with respect to the input $\bs{x}$ (pixels of the image), can significantly increase the loss and potentially lead to misclassification. Formally, the perturbation $\boldsymbol{\eta}$ is defined as:
\begin{equation}\label{3.05_fgmAttack}
  \boldsymbol{\eta}=\varepsilon\,\text{sign}\left(\nabla_xL(\boldsymbol{x},\ell)\right)
\end{equation}
where $\varepsilon > 0$ is a parameter that controls the magnitude of the perturbation. In the above expression, $\varepsilon$ represents the perturbation added per pixel to the original normalized image, where the size of the perturbation is considered over the entire image, and serves as an upper bound for the total perturbation. Adversarial examples were generated for different levels of perturbation $\varepsilon$ and using the model's own predictions as labels, to ensure that the generated examples were truly adversarial with respect to the target model.

\begin{figure}
  \centering
  \resizebox{0.5\columnwidth}{!}{\begin{tikzpicture}[x=0.9cm,y=0.9cm, font=\sffamily]

  \definecolor{scoreGreen}{RGB}{147,208,82}
  \definecolor{curvePurple}{RGB}{165,110,170}
  \definecolor{noteGreen}{RGB}{0,150,80}
  \definecolor{noteYellow}{RGB}{255,185,0}
  \definecolor{noteBlue}{RGB}{45,85,210}

  \fill[scoreGreen, fill opacity=0.20] (0,-2.0) rectangle (1.63,0.5);
  \fill[scoreGreen, fill opacity=0.20] (1.63,-2.0) rectangle (3.27,1.5);
  \fill[scoreGreen, fill opacity=0.20] (3.27,-2.0) rectangle (4.9,4.6);

  \draw[->, line width=1.1pt] (0,0) -- (10.5,0);
  \fill (0,0) circle (2.5pt);
  \node[anchor=north, scale=0.85, align=center] at (9.35,-0.1)
  {Model--human\\distance};

  \draw[densely dotted, line width=1.0pt] (1.63,-2.0) -- (1.63,5.1);
  \draw[densely dotted, line width=1.0pt] (3.27,-2.0) -- (3.27,5.1);
  \draw[densely dotted, line width=1.0pt] (4.9,-2.0)  -- (4.9,5.1);

  \node[anchor=south, scale=1.0] at (0,0.05)    {0};
  \node[anchor=south, scale=1.0] at (1.63,-0.65) {th};
  \node[anchor=south, scale=1.0] at (3.27,-0.65) {2th};
  \node[anchor=south, scale=1.0] at (4.9,-0.65)  {3th};

  \node[scale=1.4] at (0.82,-1.0) {0};
  \node[scale=1.4] at (2.45,-1.0) {1};
  \node[scale=1.4] at (4.09,-1.0) {2};

  \draw[noteGreen, line width=1.6pt, <->] (0,-1.8) -- (1.63,-1.8);
  \node[noteGreen, anchor=north, scale=1.0] at (0.82,-1.85) {th};

  \draw[curvePurple, line width=1.2pt]
  plot[smooth, domain=0:10, samples=220]
  (\x,{4.6*exp(-((\x-5.1)^2)/(2*0.80^2))});

  \draw[noteBlue, densely dotted, line width=1.6pt] (5.1,0) -- (5.1,4.75);
  \node[noteBlue, anchor=west, scale=1.1] at (5.15,0.35) {$\Delta$};

  \draw[noteYellow, line width=1.4pt, <->] (4.27,2.7) -- (5.93,2.7);
  \node[noteYellow, scale=1.3] at (6.3,2.7) {d};

\end{tikzpicture}}
  \caption{Approach used for the localization classifier. For each image, the distance $\Delta$ between the model-predicted and the human-selected coordinates is computed (blue). A Gaussian centered on $\Delta$ with standard deviation proportional to the insert diameter $d$ (yellow) is then integrated over three consecutive intervals of width $\mathit{th}$ (green), where $\mathit{th}$ is an insert-size-dependent threshold derived from the distribution of human-observer localization errors. The localization score (0, 1, or 2) is assigned as the interval yielding the largest cumulative probability.}
  \label{locClassifier}
\end{figure}
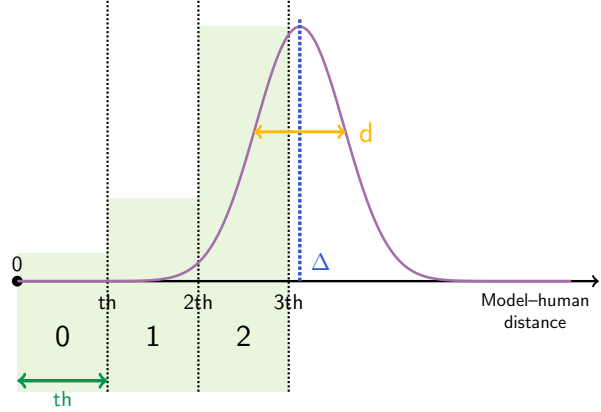

We separated the FGM attacks on the MO into two parts, in order to independently assess the robustness of the two main components of the U-Net-like architecture:
\begin{enumerate}
  \item \emph{FGM attack on the MO encoder}: the attack was applied to the MO encoder using only the confidence score prediction loss $L_{\text{MO}}$, while the localization loss was excluded. Since FGM is commonly implemented for classifiers with vectorial outputs, the scalar sigmoid confidence is mapped into a four-component representation that mimics a classifier output, with argmax corresponding to the discretization into the four classes introduced above. The perturbation is then driven by the gradient of the MSE loss $L_{\text{MO}}$ used in training, and acts on the input image so as to shift the predicted confidence score into a different class.

  \item \emph{FGM attack on the MO decoder}: attacking the MO decoder, which is responsible for the insert localization task, required the problem to be reformulated as a classification one, similarly to the score prediction mentioned above for the classification task. To this end, we implemented a \emph{localization classifier}, shown in Fig.~\ref{locClassifier}, based on the distance between human-selected coordinates and model-predicted coordinates, which outputs a score, ranging from 0 (insert correctly localized) to 2 (insert not localized). This score is computed by taking the maximum cumulative probability over three discretized intervals, for a normal distribution centered on the model predicted position relative to the human one, with a standard deviation proportional to the insert diameter. The size of the spatial intervals is defined according to specific thresholds computed from the analysis of human localization distribution for each insert diameter \cite[Supplementary Materials]{valeri2023unet}. To avoid zero-probability assignments when intervals contain only Gaussian tails (corresponding to cases where the model prediction is quite far from the human-selected position), due to numerical precision limits, a minimum probability value was imposed on each interval before the subsequent argmax step. This prevents ambiguous cases from being assigned to the default first interval (score 0) and ensures assignment to the worst-case last interval (score 2). Furthermore, when the Gaussian distribution is centered on the boundary between two intervals, symmetry produces equal probabilities for adjacent classes, and due to the argmax operation these cases are assigned to the lower-score interval. However, this is a limited issue, since the prediction is equally compatible with both classes. The resulting score is then subjected to the FGM attack, which drives the model to predict insert coordinates on adversarial images that deviate from the original predictions. In this case, the image perturbation was generated considering only the component of the loss function related to coordinate regression.
\end{enumerate}

To complement the gradient-sign attack, we also evaluated the robustness of the MO encoder using the \emph{Carlini-Wagner $L_2$} attack \cite{carlini2017towards}, a strong optimization-based adversarial method widely adopted as a benchmark for model robustness. In contrast to single-step approaches such as FGM, the CW-$L_2$ attack searches iteratively for a minimal perturbation that induces misclassification while keeping the $L_2$ norm of the perturbation as small as possible.

In our setting, the attack was applied to both the classification component, i.e.\ the encoder that predicts the confidence score, and the localization component, i.e. the decoder that predicts insert coordinates, of the MO.

Given an input CT image $\boldsymbol{x}$, the adversarial example $\boldsymbol{x}_{\text{adv}} = \boldsymbol{x} + \boldsymbol{\eta}$ is obtained by solving the optimization problem
\begin{equation}
  \label{3.06_cl2mAttack}
  \min_{\boldsymbol{\eta}}
  \left(
    \|\boldsymbol{\eta}\|_2^2 + c \cdot g(\boldsymbol{x}+\boldsymbol{\eta})
  \right)
\end{equation}
where $c>0$ balances perturbation magnitude and the misclassification objective. Following \cite{carlini2017towards}, the function $g(\cdot)$ is defined in terms of the model \emph{pseudo}-logits (discretization of the continuous encoder output) $Z(\boldsymbol{x})$ as
\begin{equation}\label{eq:g_untargeted}
  g(\bs{x}_{\text{adv}}) = \max \left( Z_\ell(\bs{x}_{\text{adv}}) - \max_{i \neq \ell} Z_i(\bs{x}_{\text{adv}}), \, -\kappa \right)
\end{equation}
where $\ell$ denotes the true class label and $\kappa \ge 0$ is the \emph{confidence parameter}.
The condition $g(\boldsymbol{x}_{\text{adv}})\le 0$ ensures misclassification, while larger values of $\kappa$ enforce a margin between the \emph{pseudo}-logits of the predicted class and that of the true class, producing higher-confidence adversarial examples at the cost of larger perturbations.

We investigated the effect of the confidence parameter $\kappa$ on the MO encoder. The attack success rate remained nearly constant up to approximately $\kappa \simeq 0.5$. This plateau is attributed to the fact that, for small $\kappa$, the adversarial examples found by the optimizer already exhibit a natural margin between the predicted and true class pseudo-logits that exceeds the specific constraint; consequently, increasing $\kappa$ within this range does not further restrict the set of admissible solutions. For $\kappa \simeq 0.5$, the required margin surpasses the pseudo-logit separation that the optimizer can achieve, and the success rate drops sharply \cite{carlini2017towards}. Based on this analysis, $\kappa$ was fixed to a representative value for all subsequent evaluations.

All adversarial examples were generated using the \emph{Adversarial Robustness Toolbox} (ART, \cite{art2018}), a Python library developed for evaluating and improving machine learning model security.  We employed the \emph{TensorFlowClassifier} estimator, which provides a standardized interface for gradient-based attacks by exposing the model's input, output, and loss tensors, along with the model definition. This configuration enables attacks such as the FGM and CW-$L_2$ to compute gradients with respect to the model input.

\subsubsection{Implementation of adversarial defense on U-Net-based model observer}\label{sectionDefense}

In the adversarial learning literature, numerous strategies have been proposed to enhance model robustness and ensure reliable inference in the presence of adversarial perturbations \cite{Chakraborty2021Survey,Pelekis2025Adversarial,Baniecki2024Adversarial}. Common defense mechanisms include input pre-processing (e.g., denoising or input transformations), information hiding (reducing model sensitivity to easily exploitable features), and model hardening (improving resistance to small perturbations)---see \cite{serban2018adversarial} for a comprehensive review. In this work, we adopt \emph{adversarial training} \cite{madry2018towards}, a well-established defense that improves robustness to small worst-case input perturbations and may also enhance generalization to CT images from different scanners and protocols.

A naive strategy of augmenting the dataset with pre-generated adversarial samples is ineffective \cite{madry2018towards}, as it only broadens input diversity without addressing the model's specific vulnerabilities. We confirmed this in preliminary experiments on our dataset by iteratively retraining the model on adversarial examples generated by the previous iteration, observing no robustness gain. Instead, following \cite{goodfellow2014explaining}, we employ \emph{dynamic adversarial training}, in which adversarial examples are generated on-the-fly during training.

In our setup, the overall loss function consists of multiple terms, as defined in
\eqref{TotalLoss_MO}. To improve robustness against attacks targeting the MO encoder,
i.e.\ the classification task, the adversarial training strategy is applied specifically to the classification loss term
$L_{\text{MO}}$. In this case, adversarial examples are generated by backpropagating
the gradients of $L_{\text{MO}}$ with respect to the input image, and the resulting
perturbed samples are used to update the network parameters while preserving the
original targets.

Formally, the classification component of the loss is replaced by
\begin{equation}
  \label{eq:adv_cls_loss}
  \bar L_{\text{MO}}
  = \alpha\, L_{\text{MO}}(\boldsymbol{x}, \ell)
  + (1-\alpha)\, L_{\text{MO}}(\boldsymbol{x}_{\text{adv}}^{\text{cls}}, \ell)
\end{equation}
where $\boldsymbol{x}_{\text{adv}}^{\text{cls}}$ denotes adversarial examples crafted
to maximize the classification error of the encoder. We set $\alpha=0.5$ to balance the contribution of clean and adversarial examples.

Analogously, to enhance robustness against attacks targeting the localization task,
adversarial training is applied to the decoder by generating adversarial examples with
respect to the localization loss $L_{\text{LOC}}$. In this case, perturbations are
computed by exploiting the gradients of $L_{\text{LOC}}$, yielding adversarial samples
$\boldsymbol{x}_{\text{adv}}^{\text{loc}}$ that aim to disrupt the predicted insert
position. The localization loss is then reformulated as
\begin{equation}
  \label{eq:adv_loc_loss}
  \bar L_{\text{LOC}}
  = \alpha\, L_{\text{LOC}}(\boldsymbol{x}, \boldsymbol{y})
  + (1-\alpha)\, L_{\text{LOC}}(\boldsymbol{x}_{\text{adv}}^{\text{loc}}, \boldsymbol{y})
\end{equation}
where $\boldsymbol{y}$ denotes the ground-truth localization targets. Again, $\alpha$ was set to $\alpha = 0.5$ to balance the contribution of clean and adversarial examples.

Depending on the experiment, adversarial training is applied either to the
classification term alone or jointly to both classification and localization terms,
while the remaining components of the total loss, including the regularization term
$L_{\text{KLD}}$, are computed on clean inputs only. This design choice allows us to
isolate and analyze the contribution of adversarial training to the robustness of each
task independently, as well as to study potential cross-task robustness effects between
classification and localization within the U-Net-based MO framework.

\subsection{Radiomic features analysis}\label{sec:radiomic_methods}

Ideally the MO should exploit the task relevant image features not affected by small, human irrelevant perturbations, to make the correct prediction and avoid to consider irrelevant features; the identification of the image features correlated to the success of the adversarial attacks may help to explain the MO behavior. In this direction we have developed a Python based code to extract and compare radiomic features from every original and adversarial image using the open-source package PyRadiomics \cite{van2017computational, zwanenburg2020ibsi}. The choice of the radiomic features has originated by two main considerations: the radiomic features represent a large, rather exhaustive, class of image features and they are also exploited in medical images to extract potential clinical information generally not detectable by human eyes.

A total of 93 features were considered, including first-order statistics
(describing intensity distributions), second-order texture features
(capturing spatial relationships between neighboring pixels) \cite{haralick1973textural},
and gray-level size zone matrix (GLSZM) descriptors \cite{Thibault2009GLSZM}. The full feature
set was selected to provide a comprehensive representation of image
characteristics potentially relevant to the model observer predictions.

The analysis was conducted on a subset of 723 original-adversarial image pairs extracted from a test dataset of 5,910 images. AEs were generated using the model trained on the first fold, with FGM ($\varepsilon = 0.001$) and CW-$L_2$ ($\kappa = 0.01$) attacks, ensuring
perturbations that are visually imperceptible while maintaining comparable
attack success rates across settings.

For each feature, the deviation from the corresponding original image was
computed and compared between successful and unsuccessful attacks. Features
were considered relevant when the difference in mean deviation between the
two groups exceeded $3\%$. This criterion enables the identification of
features associated with model vulnerability (non-robust) or stability
(robust) under adversarial perturbations.

\section{Results}

In this section, we quantify the effect of adversarial attacks on the model observer, evaluate the impact of adversarial training, and report radiomic variations induced by adversarial perturbations.

\subsection{FGM attacks on model observer} \label{results_section:FGM}

We first evaluate FGM attacks on both the encoder (classification) and decoder (localization) of the MO. Models are trained and evaluated within a 5-fold cross-validation framework: the dataset of 30,000 CT images is split into five folds---ensuring that images from the same slit are assigned to the same fold to avoid leakage bias---and five independent models are trained, each using four folds for training and one for validation. Adversarial examples were generated using the ART toolbox for multiple perturbation magnitudes $\varepsilon$.

\begin{figure}
  \centering
  \setlength{\tabcolsep}{2pt}
  \renewcommand{\arraystretch}{1.1}
  \begin{tabular}{cccc}
    \footnotesize Original
    & \footnotesize $\varepsilon = 10^{-2}$
    & \footnotesize $\varepsilon = 10^{-1}$
    & \footnotesize $\varepsilon = 1$ \\
    \includegraphics[width=0.23\columnwidth]{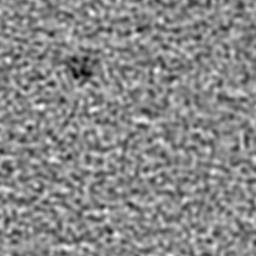} &
    \includegraphics[width=0.23\columnwidth]{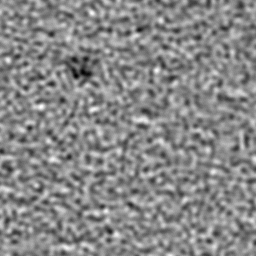} &
    \includegraphics[width=0.23\columnwidth]{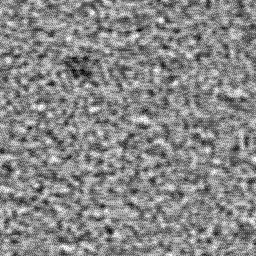} &
    \includegraphics[width=0.23\columnwidth]{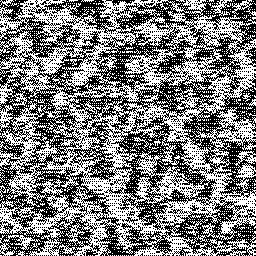} \\
    \footnotesize Score = 3
    & \footnotesize Score = 0
    & \footnotesize Score = 0
    & \footnotesize Score = 3 \\
  \end{tabular}
  \caption{FGM attacks on classification. Increasing $\varepsilon$ causes misclassification and degradation; predicted classification scores are reported below each image.}
  \label{fig3.1_InsPred}
\end{figure}

Figure~\ref{fig3.1_InsPred} shows representative FGM perturbations on the classification task. While the MO correctly identifies the target in the original image, small perturbations ($\varepsilon=10^{-2}$) are sufficient to induce misclassification despite being visually imperceptible. At larger $\varepsilon$, image quality deteriorates substantially.

\begin{figure}
  \centering
  \setlength{\tabcolsep}{2pt}
  \renewcommand{\arraystretch}{1.1}
  \begin{tabular}{cccc}
    \footnotesize Original
    & \footnotesize $\varepsilon = 10^{-2}$
    & \footnotesize $\varepsilon = 10^{-1}$
    & \footnotesize $\varepsilon = 1$ \\
    \includegraphics[width=0.23\columnwidth]{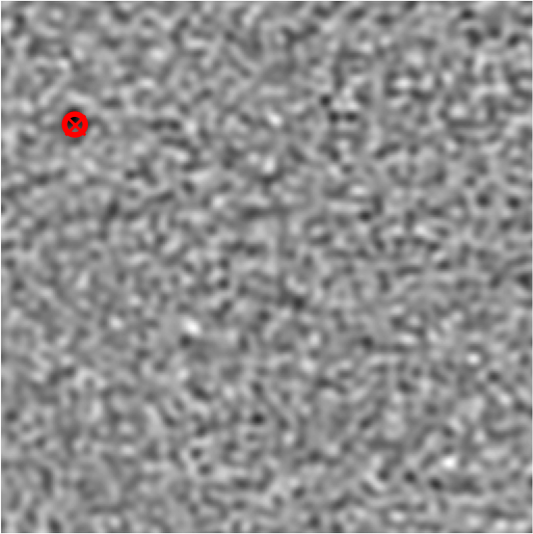} &
    \includegraphics[width=0.23\columnwidth]{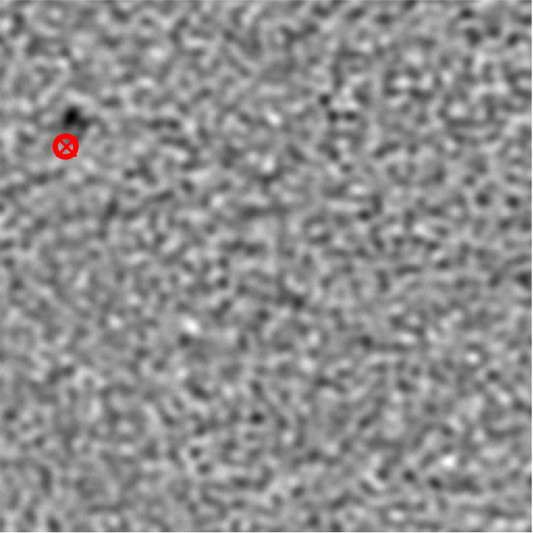} &
    \includegraphics[width=0.23\columnwidth]{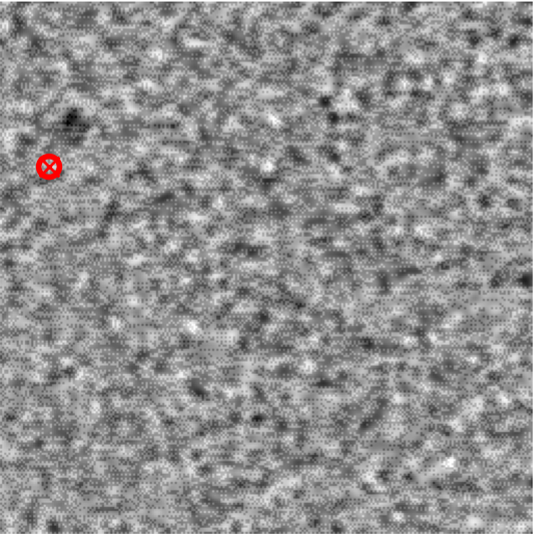} &
    \includegraphics[width=0.23\columnwidth]{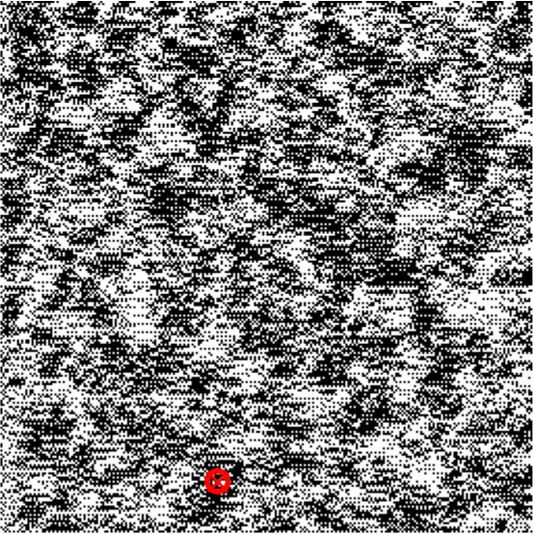} \\
    \includegraphics[width=0.23\columnwidth]{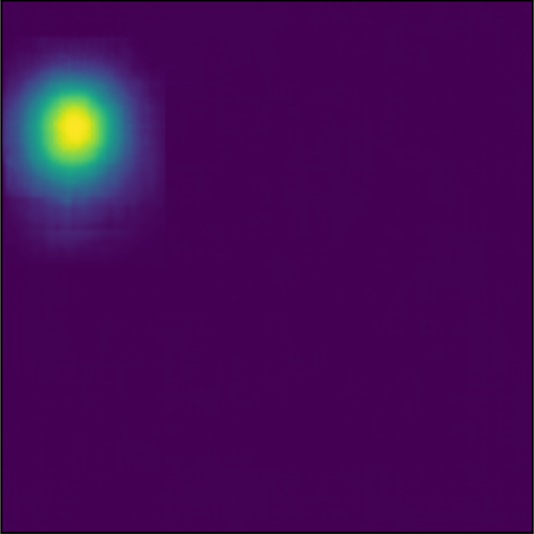} &
    \includegraphics[width=0.23\columnwidth]{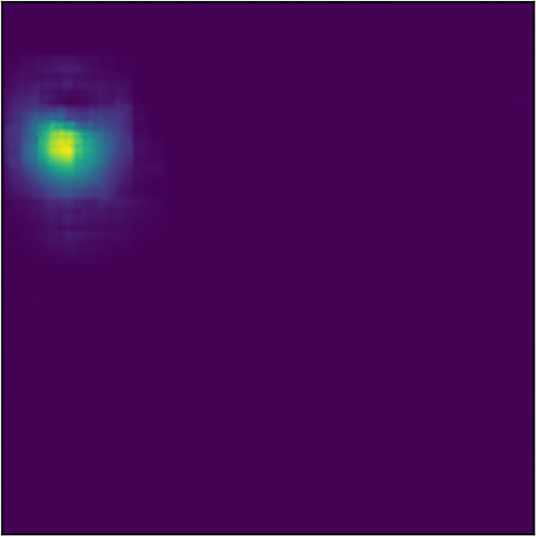} &
    \includegraphics[width=0.23\columnwidth]{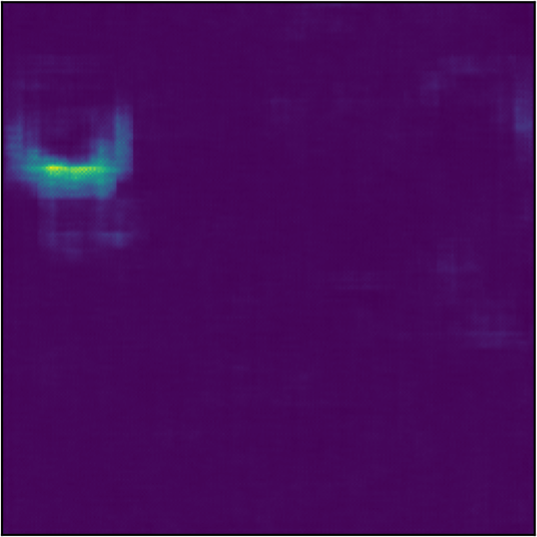} &
    \includegraphics[width=0.23\columnwidth]{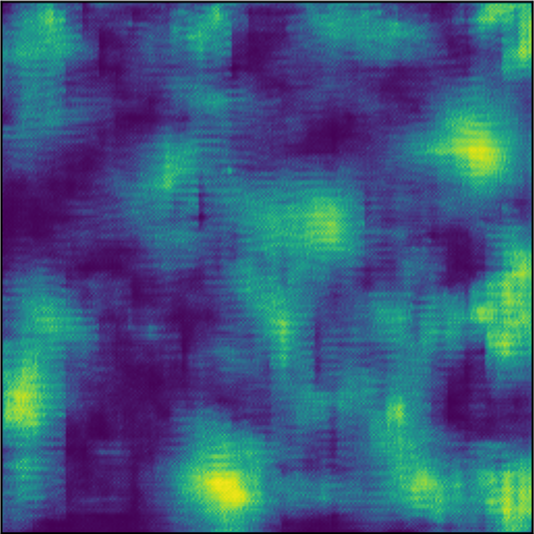} \\
    \footnotesize Localized
    & \footnotesize Not localized
    & \footnotesize Not localized
    & \footnotesize Not localized \\
  \end{tabular}
  \caption{FGM attacks on localization. Top row: adversarial input images at increasing $\varepsilon$. Bottom row: heatmaps output by the last decoder layer, whose soft-maximum yields the predicted coordinates of the low-contrast object. The localization status assigned by the MO (Localized / Not localized) is reported below each column.}
  \label{fig5.adv_examples_loc_new}
\end{figure}

A similar behavior is observed for the localization task (Fig.~\ref{fig5.adv_examples_loc_new}): as the perturbation $\varepsilon$ increases, the decoder heatmaps lose their localized peak around the insert and the predicted coordinates drift away from the true position.

\begin{figure}
  \centering
  \includegraphics[width=0.8\columnwidth]{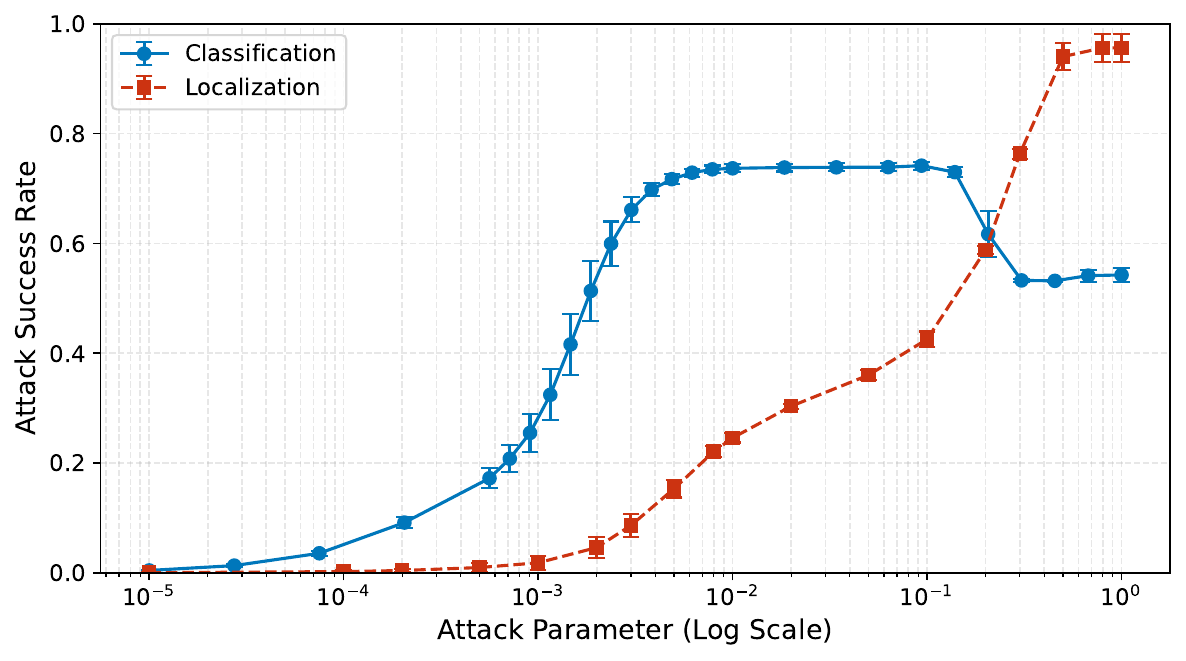}
  \caption{FGM success rates for classification and localization tasks (mean over 5 folds).}
  \label{fig6.attack_overlap}
\end{figure}

In order to quantify the adversarial effect, attack success rates were computed per fold and averaged across models. Figure~\ref{fig6.attack_overlap} summarizes classification and localization performance as a function of $\varepsilon$.

For classification, success rates increase rapidly in the intermediate regime ($\varepsilon \approx 0.002$--$0.05$), reaching up to $\sim0.75$, while remaining imperceptible. For larger perturbations ($\varepsilon \gtrsim 0.2$), performance collapses due to severe image distortion.

Localization shows lower sensitivity at small $\varepsilon$ ($<10^{-3}$), but increases up to $\sim42\%$ at moderate perturbations and approaches $100\%$ at $\varepsilon=1$.

\subsection{CW-$L_2$ attacks on model observer} \label{results_section:CWL2}

The CW-$L_2$ attack was applied to both the MO encoder (classification task) and the MO decoder (localization task). As explained in Sec.~\ref{sectionAttacks}, the confidence parameter $\kappa$ was fixed to a representative value of $\kappa = 0.01$. The attack on the MO encoder achieved a mean success rate (across the 5-folds splitting) of

\begin{equation}
  S_{CL_2M, \text{class}}^{\text{orig}} = 0.523 \pm 0.035
\end{equation}

The localization attack achieves:

\begin{equation}
  S_{CL_2M, \text{loc}}^{\text{orig}} = 0.55 \pm 0.11
\end{equation}

Compared to FGM, CW-$L_2$ generates lower success rates but produces less perceptible perturbations due to its optimization-based formulation.

\subsection{Adversarial training effects}\label{sec:advtrain}

We evaluate the effect of adversarial training (Sec.~\ref{sectionDefense}) on the robustness of the model observer against gradient-based attacks.

\begin{figure}
  \centering
  \begin{subfigure}{0.48\linewidth}
    \centering
    \caption{}
    \includegraphics[width=\linewidth]{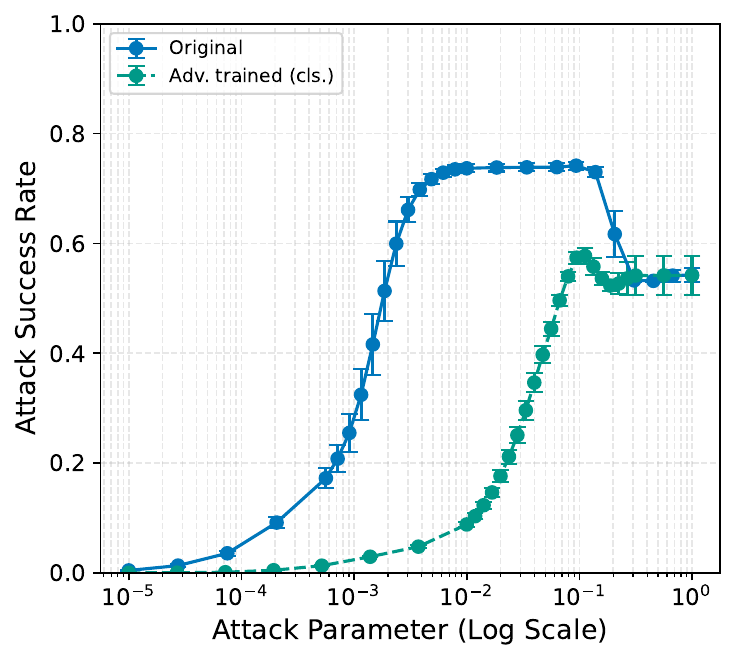}
  \end{subfigure}\hfill
  \begin{subfigure}{0.48\linewidth}
    \centering
    \caption{}
    \includegraphics[width=\linewidth]{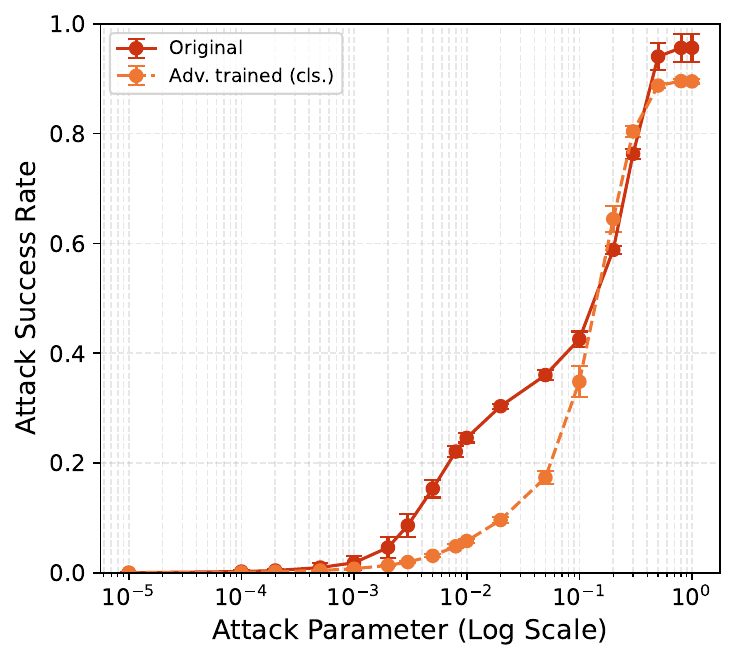}
  \end{subfigure}
  \caption{FGM attack success rates on original and classification-adversarially trained models (classification-only objective) for (a) classification and (b) localization. Mean $\pm$ standard error of the mean (SEM) over 5-fold cross-validation; logarithmic $\varepsilon$ scale.}
  \label{fig7.advtrain1}
\end{figure}

Figure~\ref{fig7.advtrain1} shows the FGM attack effect on the model trained dynamically with adversarial examples specifically crafted to fool the classification task, compared to the original MO. Classification attack success rate in Fig.~\ref{fig7.advtrain1}a shows that for small perturbations ($\varepsilon \lesssim 0.01$), adversarial training provides near-complete robustness, with almost zero attack success. In the intermediate regime, the success rate increases but remains consistently lower than the original model, peaking at $\sim 0.6$ around $\varepsilon = 0.1$. For large perturbations ($\varepsilon \gtrsim 0.2$), both models converge to similar behavior, as strong distortions dominate the input distribution.

Interestingly, as shown in Fig.~\ref{fig7.advtrain1}b, classification-only objective adversarial training improves not only classification robustness but also localization performance, particularly in the low-perturbation regime ($\varepsilon=10^{-3}$--$10^{-1}$), suggesting more general robustness in shared feature representations.

\begin{figure}
  \centering
  \begin{subfigure}{0.48\linewidth}
    \centering
    \caption{}
    \includegraphics[width=\linewidth]{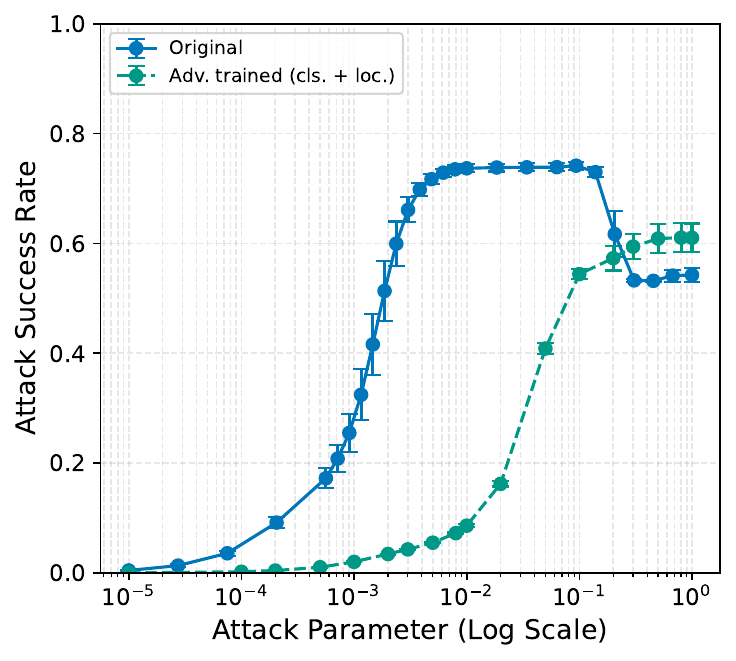}
  \end{subfigure}\hfill
  \begin{subfigure}{0.48\linewidth}
    \centering
    \caption{}
    \includegraphics[width=\linewidth]{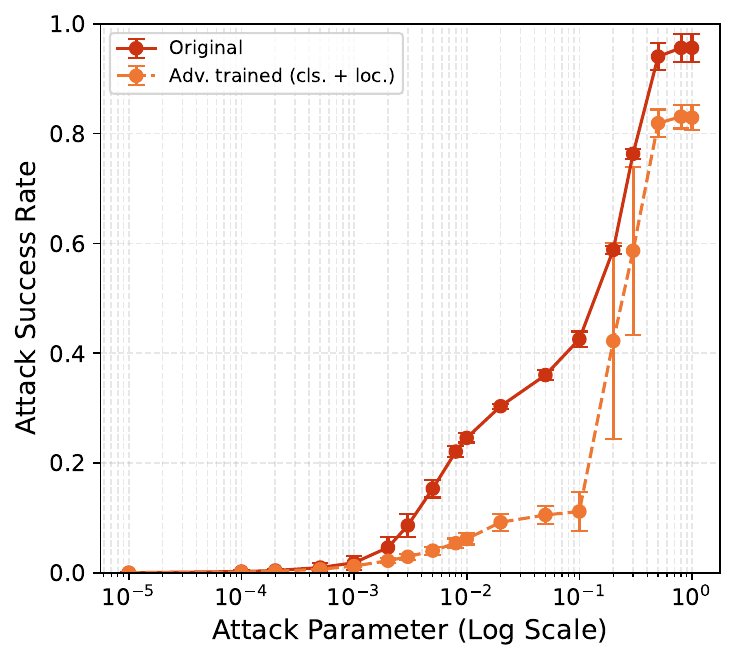}
  \end{subfigure}
  \caption{FGM attack success rates for original and jointly adversarially trained models for (a) classification and (b) localization. Mean $\pm$ SEM over 5-fold cross-validation; logarithmic $\varepsilon$ scale.}
  \label{fig.overlap_retrain}
\end{figure}

Extending training with a localization-specific adversarial loss further reduces attack success on localization while preserving classification robustness, as shown in Fig.~\ref{fig.overlap_retrain}

We further evaluate robustness using the CW-$L_2$ attack. For classification-only adversarial training, the attack success rate drops to
\begin{equation}
  S_{CL_2M}^{\text{adv-class}} = 0.0726 \pm 0.0031
\end{equation}
showing a strong suppression of gradient-based optimization attacks.

Including localization-aware adversarial loss yields
\begin{equation}
  S_{CL_2M}^{\text{adv-loc}} = 0.125 \pm 0.017
\end{equation}
which remains significantly lower than the original model and comparable to the classification-only case. Overall, both strategies confirm that adversarial training substantially improves robustness of the encoder and generalizes across tasks.

\subsection{Effect on model performance}\label{sec:effect_perf}

We next assess whether the robustness gains achieved through adversarial training affect the primary task performance of the model observer, namely CT image quality assessment for protocol optimization.

\begin{figure}
  \centering
  \includegraphics[width=0.8\linewidth]{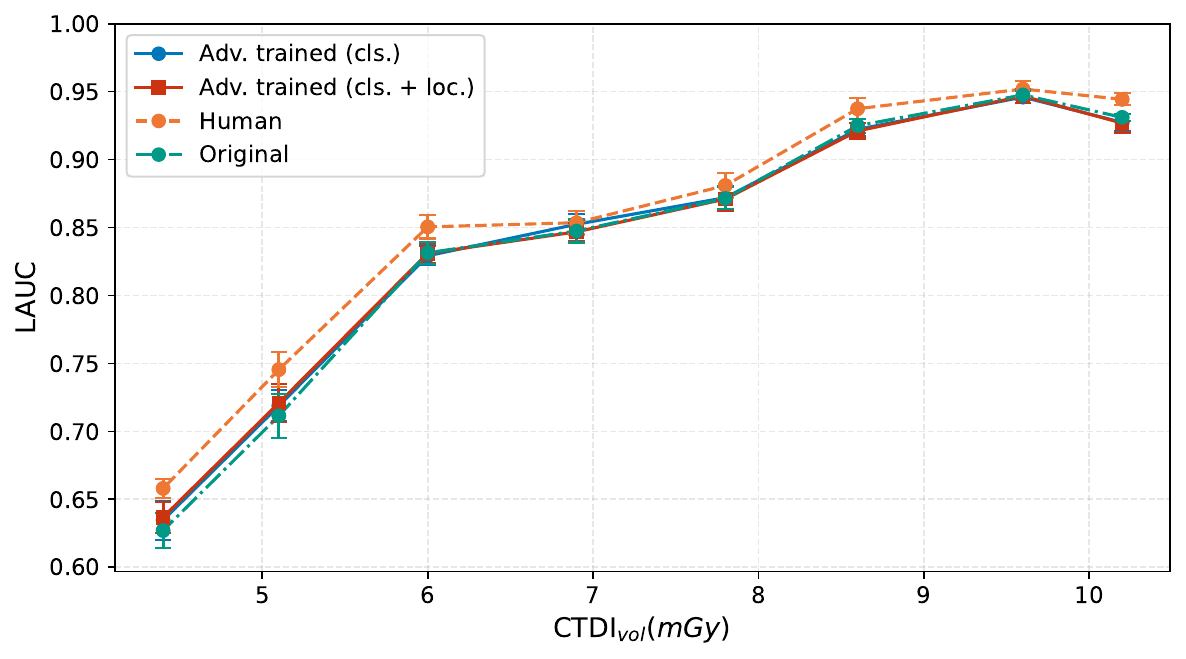}
  \caption{LROC curves comparing human observers with the original and adversarially trained models. Results are averaged over a 5-fold cross-validation.}
  \label{fig4.4_lroc-train}
\end{figure}

Diagnostic performance of the original and adversarially trained models is evaluated against human observers using localization receiver operating characteristic (LROC) analysis and the corresponding area under the curve (LAUC) \cite{swensson1996lroc}, shown in Fig.~\ref{fig4.4_lroc-train} as a function of CTDI. Results are reported as mean curves over a 5-fold cross-validation.

The LROC curves of the adversarially trained model closely match those of the original network and remain consistent with human performance. Correspondingly, LAUC values show no statistically significant difference after adversarial training, indicating preservation of both detection sensitivity and localization accuracy.

Overall, these results indicate that the robustness improvements reported in the previous section are achieved without degradation of task-specific performance.

\subsection{Radiomic characterization of adversarial perturbations}\label{sec:radiomic_results}

Radiomic analysis of the 93 extracted features shows that most descriptors exhibit negligible variation between original and adversarial images, with mean deviations close to zero for both successful and unsuccessful attacks. Overall, global intensity statistics are largely preserved under adversarial perturbations.

\begin{figure}
  \centering
  \begin{subfigure}{0.48\linewidth}
    \centering
    \caption{}\label{fig:radiomicsFGMClass}
    \includegraphics[width=\linewidth]{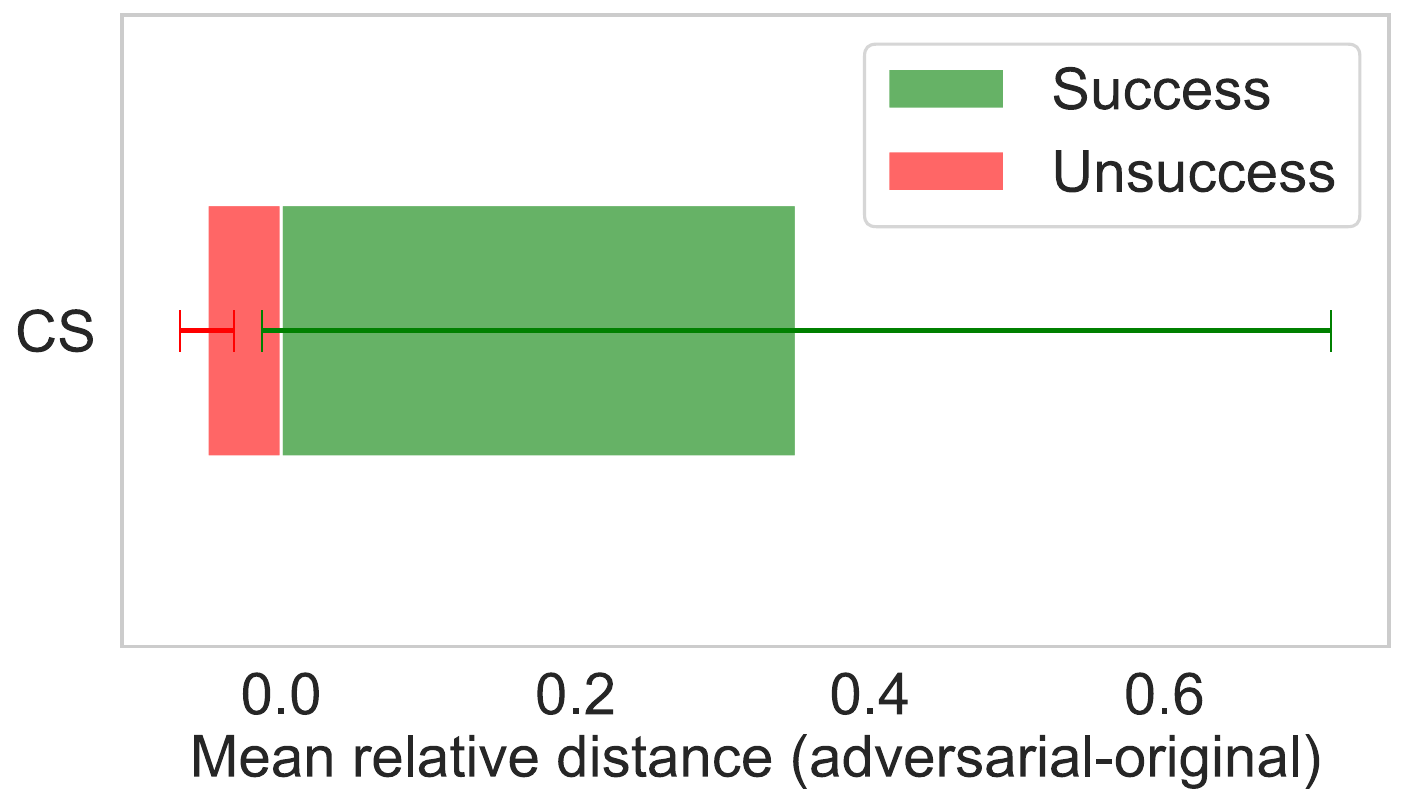}
  \end{subfigure}\hfill
  \begin{subfigure}{0.48\linewidth}
    \centering
    \caption{}\label{fig:radiomicsFGMLoc}
    \includegraphics[width=\linewidth]{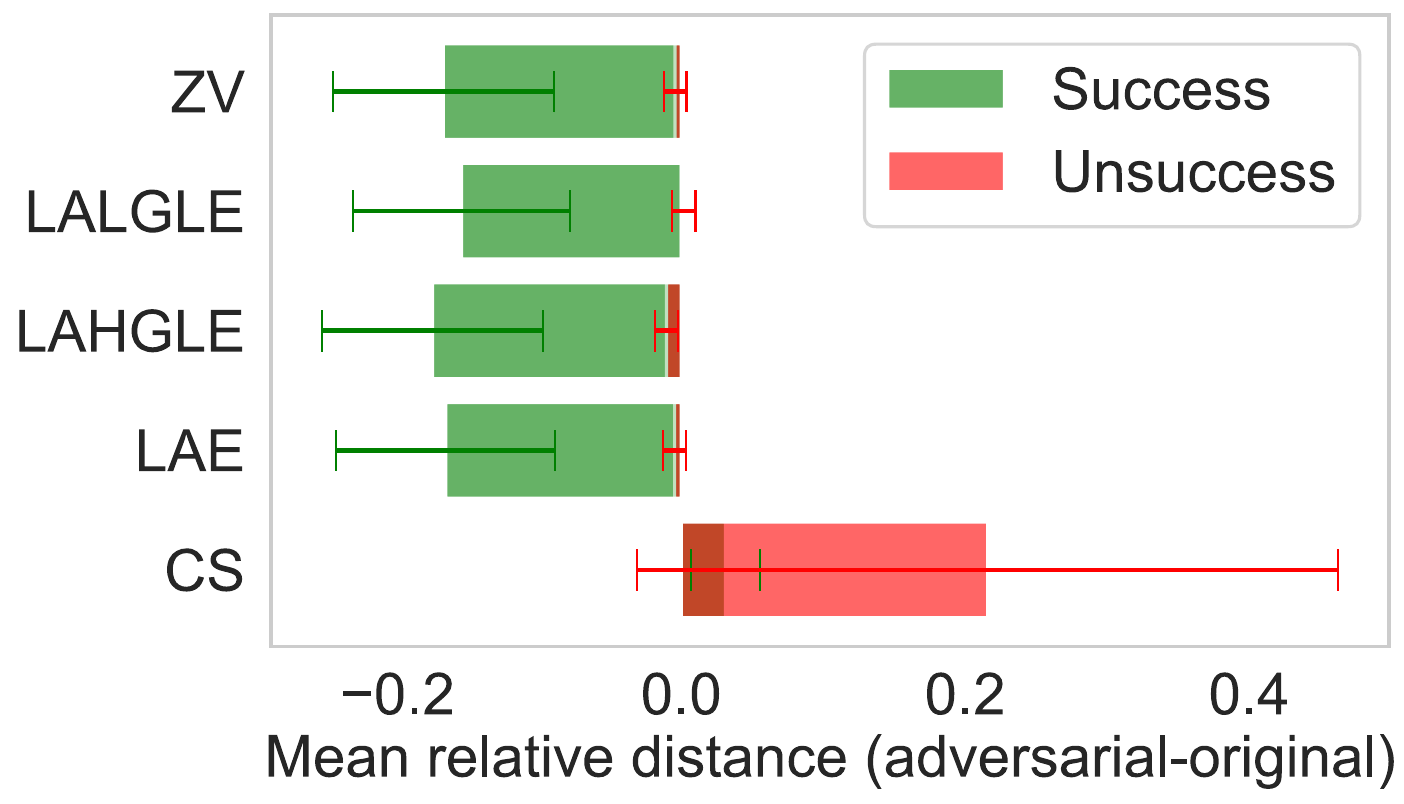}
  \end{subfigure}
  \caption{Mean relative variation (adversarial-original) of selected radiomic features for FGM attacks ($\varepsilon = 0.001$): (a) classification and (b) localization. CS = \emph{ClusterShade}, ZV = \emph{ZoneVariance}, LALGLE = \emph{LargeAreaLowGrayLevelEmphasis}, LAHGLE = \emph{LargeAreaHighGrayLevelEmphasis}, LAE = \emph{LargeAreaEmphasis}.}
  \label{fig.radiomicsFGM}
\end{figure}

\begin{figure}
  \centering
  \begin{subfigure}{0.48\linewidth}
    \centering
    \caption{}\label{fig:radiomicsCLClass}
    \includegraphics[width=\linewidth]{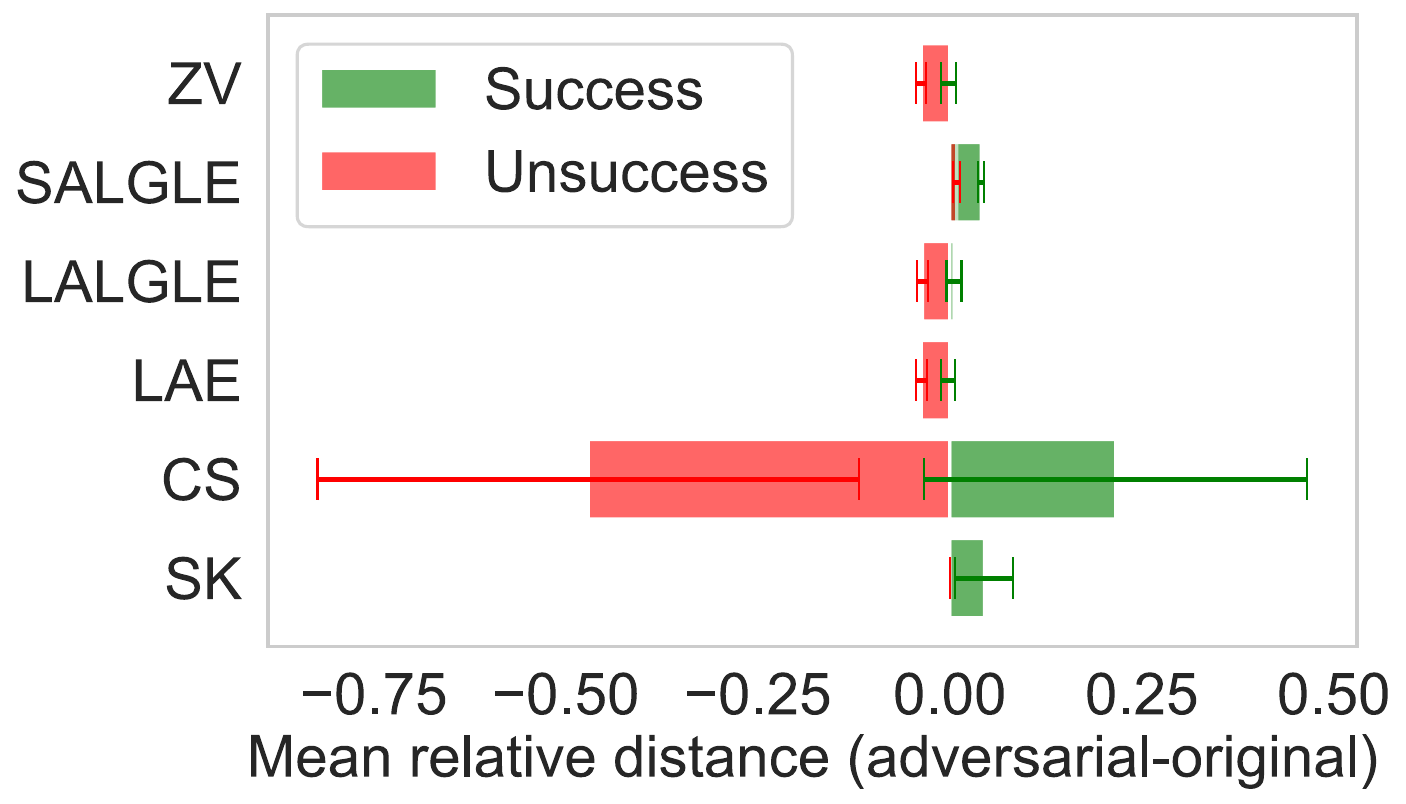}
  \end{subfigure}\hfill
  \begin{subfigure}{0.48\linewidth}
    \centering
    \caption{}\label{fig:radiomicsCLLoc}
    \includegraphics[width=\linewidth]{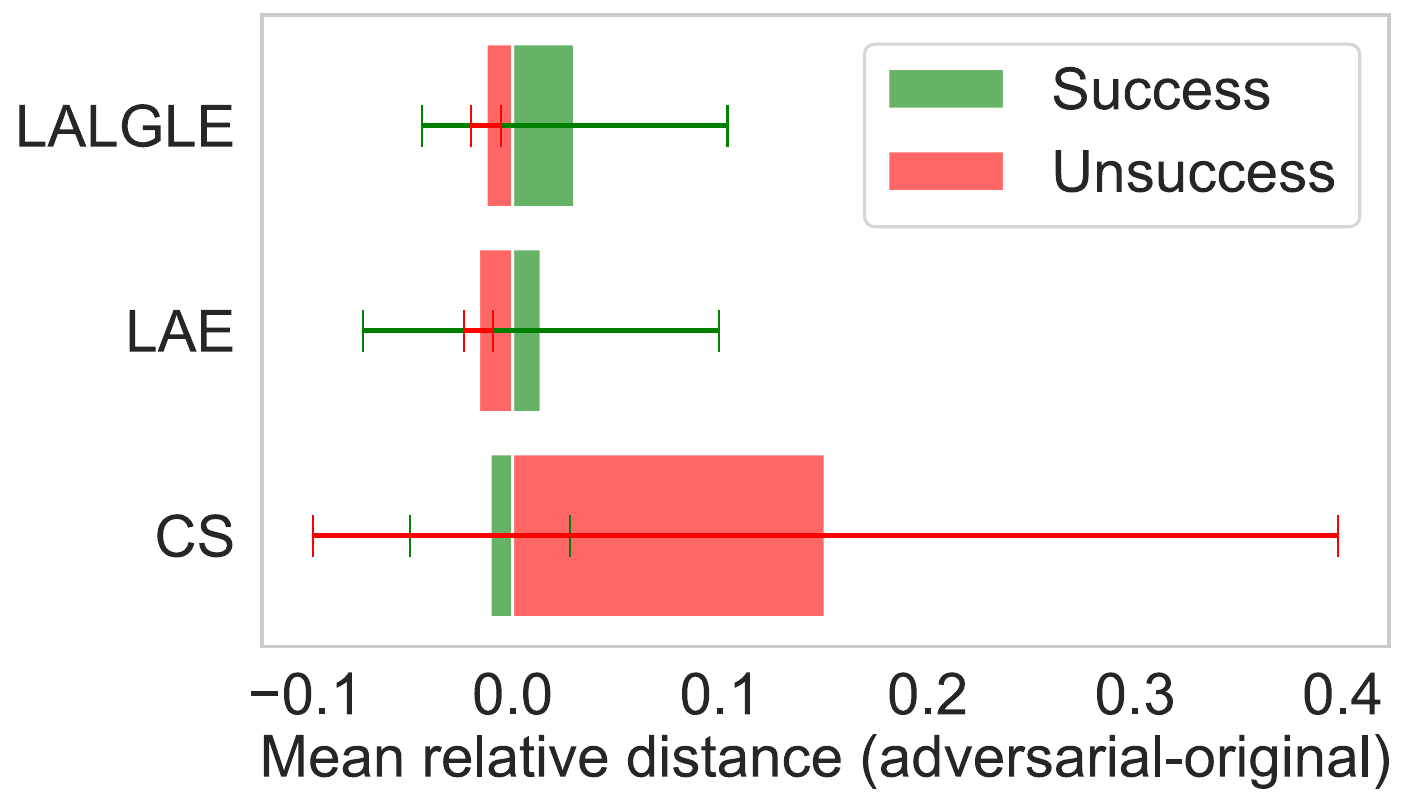}
  \end{subfigure}
  \caption{Mean relative variation (adversarial-original) of selected radiomic features for CW-$L_2$ attacks ($\kappa = 0.01$): (a) classification and (b) localization. CS = \emph{ClusterShade}, SK = \emph{Skewness}, ZV = \emph{ZoneVariance}, LALGLE = \emph{LargeAreaLowGrayLevelEmphasis}, SALGLE = \emph{SmallAreaLowGrayLevelEmphasis}, LAE = \emph{LargeAreaEmphasis}.}
  \label{fig.radiomicsCL}
\end{figure}

However, a subset of features shows consistent differences between successful and unsuccessful attacks, mainly among second-order texture descriptors. The observed feature distributions for FGM and CW-$L_2$ attacks are reported in Fig.~\ref{fig.radiomicsFGM} and Fig.~\ref{fig.radiomicsCL} respectively.

In particular, \emph{ClusterShade} is the most sensitive feature across all settings. For classification attacks, successful perturbations induce a positive shift of approximately +35\% for FGM and +20\% for CW-$L_2$ (Fig.~\ref{fig:radiomicsFGMClass} and Fig.~\ref{fig:radiomicsCLClass}, respectively). In contrast, localization attacks show minimal variation in the same feature (Fig.~\ref{fig:radiomicsFGMLoc} and Fig.~\ref{fig:radiomicsCLLoc}).

For CW-$L_2$ classification attacks, the first-order feature \emph{Skewness} shows a moderate increase (approximately 5\%) (Fig.~\ref{fig:radiomicsCLClass}).

Additional changes are observed in GLSZM-based features: FGM localization attacks produce a negative shift of approximately $-17\%$ (Fig.~\ref{fig:radiomicsFGMLoc}), while CW-$L_2$ induces weaker and less consistent variations. For CW-$L_2$ classification attacks, a small positive shift of approximately 4\% is observed in one GLSZM descriptor (Fig.~\ref{fig:radiomicsCLClass}).

Overall, radiomic variations are concentrated in second-order texture features, altering spatial intensity relationships while preserving global image and first-order intensity statistics.

\section{Discussion}

CNN-based model observers are highly effective for CT image quality assessment but remain vulnerable to adversarial perturbations, which can compromise both classification and localization performance under clinically realistic conditions.

Our results demonstrate that FGM and CW-$L_2$ attacks expose distinct failure modes: FGM efficiently identifies decision boundaries via gradient steps, while CW-$L_2$ produces smaller but more controlled perturbations at higher computational cost.

A consistent observation is that localization is more robust than classification at low perturbation levels (Fig.~\ref{fig6.attack_overlap}), likely due to the spatial inductive bias of the U-Net architecture. However, when comparing attacks across methods, CW-$L_2$ achieves comparable localization success rates at lower and less perceptible perturbation magnitudes than FGM, which typically requires larger, visually noticeable distortions. This indicates that optimization-based attacks are more efficient in generating subtle yet effective perturbations for spatial prediction tasks.

Adversarial training substantially improves the robustness of the model observer against gradient-based attacks (Fig.~\ref{fig7.advtrain1} and Fig.~\ref{fig.overlap_retrain}). In particular, training with adversarial examples crafted for the classification task not only increases resistance to classification attacks, but also enhances resilience to localization perturbations, especially in the low-perturbation regime ($\varepsilon = 0.001$--$0.1$), where distortions remain imperceptible (Fig.~\ref{fig7.advtrain1}). This indicates that adversarial training promotes a more general robustness in the shared feature representation learned by the network.

This improved robustness is further confirmed when considering the CW-$L_2$ attack. The marked reduction in attack success rate demonstrates that adversarial training effectively suppresses the optimizer's ability to identify crafted perturbations that induce misclassification. Moreover, extending adversarial training to include both classification and localization objectives yields comparable suppression of CW-$L_2$ attacks, indicating that robustness generalizes across tasks and confirming the stability of the encoder under both defense strategies.

The ROC analysis in Fig.~\ref{fig4.4_lroc-train} highlights a crucial finding: while adversarial training improves robustness to perturbations and adversarial noise, it does not compromise the model's interpretability or its task-specific performance. A trade-off between adversarial robustness and standard accuracy has been reported in the literature \cite{tsipras2019robustness, zhang2019theoretically}, although its extent is known to depend on the model architecture, the training regime, and the specific task. The stability of the LAUC metric in our setting suggests that, under the conditions examined here, robustness and task effectiveness can coexist when adversarial defenses are properly integrated into the training pipeline.

Radiomic analysis shows that adversarial vulnerability is primarily associated with specific second-order texture features rather than global intensity statistics. In particular, variations in ClusterShade indicate that changes in local intensity relationships are key to misleading the encoder (Fig.~\ref{fig.radiomicsFGM}). In contrast, localization failures show minimal variation in this feature, suggesting lower sensitivity of spatial prediction to such local texture perturbations.

For CW-$L_2$ attacks, additional effects are observed in first-order statistics (Fig.~\ref{fig.radiomicsCL}), where shifts in Skewness suggest that global intensity asymmetry may contribute to misclassification. Moreover, features derived from GLSZM, which describe the spatial distribution of homogeneous intensity regions, show consistent variations, indicating that reorganization of spatial intensity structure also contributes to prediction failures.

Overall, these findings suggest that the model observer is particularly sensitive to subtle modifications of local texture patterns, which are not perceptible to human observers but are sufficient to alter model predictions. Radiomic analysis therefore provides a means to distinguish between non-robust features, which vary significantly between successful and unsuccessful attacks and likely drive misclassification, and robust features, which remain stable and are less associated with prediction changes. This distinction offers an interpretable framework for understanding model vulnerabilities and can guide the design of model observers that rely more strongly on stable image descriptors, improving robustness to small perturbations.

\section{Conclusion}

CNN-based models achieve strong performance in medical imaging but remain sensitive to small adversarial perturbations, raising concerns for safety-critical applications such as clinical decision support and CT protocol optimization under the ALARA principle. In this study, we evaluate adversarial training to improve the robustness of a U-Net-based model observer for CT image quality assessment and low-contrast object detection and localization.

Adversarial perturbations are used to systematically probe model sensitivity to targeted input changes. After adversarial training, the model shows substantially increased robustness to both gradient-based and optimization-based attacks, while preserving its primary task performance. This is confirmed by stable LROC curves and unchanged area under the LROC metric compared to the original model and human observers, indicating no degradation in clinical utility.

These results demonstrate that robustness and task performance can be jointly achieved when adversarial defenses are properly integrated into training. More broadly, adversarial training supports the development of more reliable and generalizable medical imaging AI systems capable of maintaining performance under realistic data variability, an important requirement for clinical translation and regulatory standards.

Additionally, adversarial perturbations provide insight into model behavior by revealing sensitive input regions, while radiomic analysis helps identify texture features associated with failure modes, supporting improved interpretability and feature robustness \cite{welch2019vulnerabilities}. Future work will extend these findings to other architectures and investigate explainability methods, including activation mapping and feature attribution \cite{selvaraju2017gradcam}, to better understand how adversarial perturbations affect internal representations and decision mechanisms.


\section*{Declaration of competing interest}
The authors declare that they have no known competing financial
interests or personal relationships that could have appeared to influence
the work reported in this paper.

\section*{Acknowledgments}
This work was carried out within the framework of an inter-institutional collaboration agreement involving Azienda Ospedaliero-Universitaria Careggi, Department of Physics and Astronomy of the   Universit\`{a} degli Studi di Firenze, Istituto Superiore di Sanit\`{a}, Azienda USL Toscana Centro, Fondazione Bruno Kessler, National Research Council of Italy, and UNISER Pistoia. The authors acknowledge the financial support provided by Careggi and AUSL Toscana Centro through research fellowships, as well as the computational infrastructure made available by UNISER Pistoia and by Istituto Superiore di Sanità as a partner of Project ECS 0000024 Rome Technopole, – CUP B83C22002820006, NRP Mission 4 Component 2 Investment 1.5, funded by the European Union – NextGenerationEU. The contribution of all researchers and professionals involved in the collaboration is gratefully acknowledged, with particular recognition of the interdisciplinary exchanges and staff mobility across the participating institutions, which enabled the sharing of expertise and the advancement of the different research lines leading to these results.

\section*{Data Availability}
The datasets generated and/or analyzed during the current study are not publicly available due to ongoing research activities but are available from the corresponding author on reasonable request.

\bibliographystyle{plain}
\bibliography{references}

\end{document}